# Temperature and Competition: Drivers in the Ecological Dynamics of *Aedes* Mosquitoes and Dengue Spread

Santiago Andrés Villamil-Chacón[1,2*] and Mauricio Santos-Vega[1,2*]

[1*]School of Life Sciences, Universidad de los Andes, Cra 1 № 18A - 12, Bogotá D.C, 111711, Cundinamarca, Colombia.
[2]Grupo Biología Matemática y Computacional (BIOMAC), Universidad de los Andes, Cra 1 № 18A - 12, Bogotá D.C, 111711, Cundinamarca, Colombia.

*Corresponding author(s). E-mail(s): sa.villamil@uniandes.edu.co; om.santos@uniandes.edu.co;

## Abstract

**Background:** Dengue is a mosquito-borne viral disease endemic to tropical regions, primarily transmitted by *Aedes aegypti* and *Aedes albopictus*. Climate-driven temperature changes are altering vector ecology and expanding the geographic range where both species coexist. However, the combined effects of temperature variability and interspecific interactions—particularly the highly competitive larval stage—on mosquito population dynamics and dengue transmission remain poorly understood.

**Methods:** We developed a deterministic model incorporating temperature-dependent parameters to analyze vector interactions across larval stage, coupled with a SEIR framework for human infection dynamics. We evaluated species invasion capability, population dynamics, and transmission patterns through invasion and coexistence analyses, as well as infection peak assessment. The basic reproduction number ($R_0$) was derived analytically using the Next Generation Matrix (NGM) method, while the effective reproduction number ($R_t$) was computed from numerical simulations to capture dynamic effects of larval competition.

**Results:** The invasion analysis showed that *Ae. albopictus* successfully invaded *Ae. aegypti*-dominated systems under temperature-independent conditions when *Ae. aegypti*'s larval competition coefficient ($\omega_{ae}$) was below 0.47, with neutral equilibrium at 0.47-0.60 and exclusion above 0.60. In temperature-dependent conditions, invasion potential expanded, indicating positive Lyapunov coefficients up to $\omega_{ae}$ = 0.75. Coexistence analysis revealed *Ae. aegypti* dominance (87.5% mean relative abundance) in temperature-independent scenarios, while temperature-dependent scenarios enabled balanced coexistence, with both species near 50% mean relative abundance. Dengue transmission peaked at 156-168 infected individuals under temperature-independent conditions and 195-220 under temperature-dependent conditions, with higher *Ae. albopictus* competition reducing infection peaks in both scenarios. The basic reproduction number ($R_0$) was indirectly influenced by larval competition via adult mosquito abundance, with $R_t$ ranging from 2.05-2.20 in temperature-independent scenarios and 1.00-1.80 in temperature-dependent scenarios with increased larval competition.

**Conclusions:** Temperature significantly influences competitive interactions and dengue transmission dynamics between *Ae. aegypti* and *Ae. albopictus*. Temperature-dependent conditions enhance *Ae. albopictus* invasion and promote coexistence, while *Ae. aegypti* drives higher infection peaks under favorable thermal conditions. Increased Ae. albopictus competition lowers transmission, particularly in temperature-dependent scenarios; however, situations where both vectors exhibit similar abundances represent the most concerning context. These findings underscore the importance of integrating temperature effects and interspecific competition into vector control strategies in regions like Colombia, where both species coexist, to effectively mitigate dengue transmission under varying climatic conditions.

# 1 Introduction

Arboviruses, a group of viruses transmitted primarily by hematophagous mosquitoes, represent a major global public health concern [1]. Among these, members of the *Flaviviridae* family—particularly dengue virus (DENV) and Zika virus (ZIKV)—are of critical importance due to their prevalence in tropical and subtropical regions, where environmental conditions support the growth of mosquito populations [2]. Dengue, in particular, has reached alarming incidence levels. In the Americas, the Pan American Health Organization (PAHO) reported 12.8 million suspected dengue cases in 2024 [3]. Colombia is one of the most affected countries, with dengue representing the predominant flavivirus, which circulates mainly in warm, humid regions such as coastal areas and Andean valleys. In 2024, the country recorded 314,915 cases—equivalent to 944.2 per 100,000 inhabitants—marking a steep increase from 346.8 per 100,000 in 2023 [4].

In Colombia, dengue transmission is primarily driven by *Aedes aegypti*, which is adapted to urban environments. Although *Aedes albopictus* is reported less frequently, it has the potential to act as a secondary vector [5]. Both species are generally found at elevations below 2,200 meters above sea level, yet they differ markedly in ecological preferences and vector competence. *Ae. aegypti* thrives in warmer urban environments (24–34°C) and breeds predominantly in artificial containers, whereas *Ae. albopictus* is more prevalent in cooler suburban and rural environments (15–30°C), and breeds in natural water reservoirs [6]. Although both species have similar life cycles, their feeding behaviors differ significantly. *Ae. aegypti* shows a strong preference for human hosts, feeding on human blood 95% of the time, and exhibits bimodal biting activity. On the other hand, *Ae. albopictus* has a more opportunistic feeding pattern, with a 60% preference for humans, and tends to feed primarily during the day [7–9]. These differences in feeding behavior, along with reproductive strategies, such as *Ae. aegypti* feeding on blood every 3 to 5 days [10], contribute to its greater efficiency as a dengue vector, despite *Ae. albopictus* having a wider range of potential hosts [11, 12].

Both species undergo the same aquatic developmental stages: egg, larva, and pupa. During the larval phase, individuals are confined to limited water resources, making this stage the most competitive in their life cycle. At this time, larvae directly compete for space and food. The survival of individuals during this phase is crucial, as it significantly impacts adult population size and, consequently, transmission potential [13]. Interspecific competition among larvae can hinder the population growth of one species, alter dominance patterns, and change the composition of adult vector communities, ultimately influencing the epidemiology of dengue.

Climate change has contributed to the expansion of *Ae. albopictus* into temperate regions, including areas that are already home to *Ae. aegypti* [14]. Human activities, such as global trade, further facilitate this expansion by creating new breeding sites for *Ae. albopictus* [15]. The ecological adaptability of *Ae. albopictus* enables it to thrive in a wide range of climates, from cooler rural areas to warmer peri-urban environments [16]. In contrast, *Ae. aegypti* is primarily found in densely populated urban spaces. As a result, *Ae. albopictus* has become the main vector in some non-urban regions, leading to outbreaks in places like China, Mauritius, and Hawaii [15, 17–19].

Recent reports indicate that the mosquito species *Ae. aegypti* and *Ae. albopictus* are co-occurring in localities in central Colombia, such as Ibagué (Tolima) [20]. This situation emphasizes the urgent need to understand how their interactions may influence the epidemiology of dengue. In areas where both species overlap, competition during their larval stage, considered the most competitive phase of their life cycle, could play a critical role in determining which species will dominate and how dengue transmission occurs. These coexistence scenarios create complex ecological and epidemiological dynamics, especially under temperature regimes altered by climate change.

To address this, we develop a temperature-dependent deterministic model that examines how thermal variation and larval-stage competition between *Ae. aegypti* and *Ae. albopictus* influence dengue transmission in coexistence contexts. Our approach integrates temperature-dependent mosquito traits with explicit modeling of interspecific competition in the larval stage, enabling us to assess how temperature variations may facilitate *Ae. albopictus* invasion into *Ae. aegypti*-dominated areas. We hypothesize that while *Ae. aegypti*'s stronger anthropophilic feeding will sustain higher transmission potential in most scenarios, temperature-driven larval competition will generally reduce overall vector abundance, thereby decreasing dengue cases. This suppression effect is expected to be strongest when competition between species is bidirectional.

# 2 Methods

## 2.1 Model description

We developed a deterministic model to analyze how temperature-dependent competition between the larval stages of *Ae. aegypti* and *Ae. albopictus* affects dengue transmission dynamics (Figure 1). The model is based on several key assumptions. First we consider only one dengue serotype, which eliminates the possibility of reinfection. Additionally, we focus exclusively on female mosquitoes, as they are crucial for disease transmission due to their blood-feeding behavior on human hosts. For each vector species (*Ae. aegypti* and *Ae. albopictus*), we defined three compartments: one for larvae, one for susceptible adult females, and one for infected adult females, with infection occurring through contact between infected humans and susceptible mosquitoes. For the human population, we employ an SEIR model structure, where individuals progress from susceptible to exposed (via mosquito bite), then infected when symptoms manifest, and finally recovered with full immunity to the virus serotype.

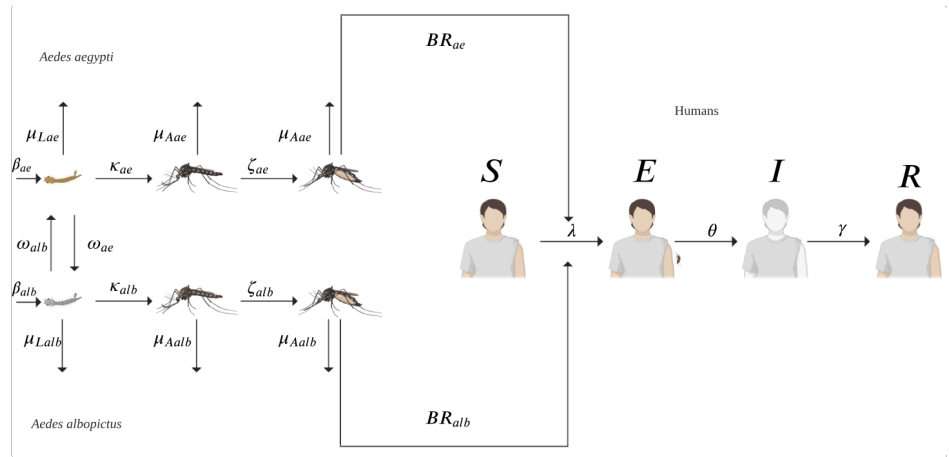

**Fig. 1**: Schematic of the temperature-dependent dengue transmission model. The compartmental diagram captures coupled vector-human dynamics, with two *Aedes* species (*Ae. aegypti* and *Ae. albopictus*) each structured into larval, susceptible adult female, and infected adult female stages. Larval populations compete within habitats shaped by temperature, and their development into adults depends on thermal conditions. Adult females acquire dengue by biting infected humans (SEIR compartments: Susceptible → Exposed → Infected → Recovered), then transmit the virus to susceptible humans. The model assumes: (1) exclusive female mosquito vectoring, (2) single serotype transmission (no reinfection), and (3) temperature-driven larval competition regulating adult population sizes. Solid arrows denote transition pathways influenced by temperature, species interaction, and dengue infection status.

We modified a baseline temperature-independent model by incorporating temperature-dependent parameters, represented as functions of the form $f(g(t))$. This approach allows us to model key mosquito life-history traits—including development rate, mortality, transmission probability, and biting rate—based on functional approximations found in the literature [21–28]. For site-specific simulations, we implemented a temperature function $g(t)$ that replicates the annual thermal regime of Ibagué (25-31°C; Table 1). This dynamic temperature range effectively captures the overlapping thermal tolerances of both *Ae. aegypti* and *Ae. albopictus*, which allow us to analyze their competitive interactions under realistic climate variability. It also provides a potential analogue for other Colombian localities that may be affected by climate change.

We conducted three analyses for both temperature-dependent and temperature-independent models. First, we performed an invasion analysis in a deterministic environment to evaluate whether *Ae. albopictus* could successfully invade areas dominated by established *Ae. aegypti* populations. Second, we conducted a coexistence analysis to determine if *Ae. albopictus* populations could persist in the long term, beyond temporary invasion, while facing competitive pressure from *Ae. aegypti*. Finally, we examined disease dynamics to assess how these ecological interactions influence dengue transmission. This approach distinguishes between transient invasion events and stable coexistence, while also quantifying their epidemiological consequences.

### 2.1.1 Mathematical model

This study presents a mathematical model that describes the interactions between the populations of the two mosquito vector species, *Ae. aegypti* (*ae*) and *Ae. albopictus* (*alb*), along with the transmission dynamics of the dengue virus to humans. The model consists of three main components: larval dynamics, adult mosquito dynamics, and human epidemiological dynamics. Each equation is designed to incorporate speciesspecific biological parameters as well as interactions between the two mosquito species.

The larval dynamics for each mosquito species ($L_{ae}$ and $L_{alb}$) are influenced by larval reproduction rates ($\beta_{ae}$ and $\beta_{alb}$), larval carrying capacity ($K$), and larval competition coefficients ($\omega_{ae}$ for *Ae. aegypti* affecting *Ae. albopictus*, and $\omega_{alb}$ for the reverse interaction). The rates $\beta_{ae}$ and $\beta_{alb}$ indicate the addition of larvae to the population, while the larval development rates ($\kappa_{ae}$ and $\kappa_{alb}$) describe the transition from larvae to adult mosquitoes. Larval mortality rates ($\mu_{Lae}$ and $\mu_{Lalb}$) account for natural mortality during this life stage. Uninfected adult mosquitoes ($X_{ae}$ and $X_{alb}$) emerged from larvae at rates $\kappa_{ae}$ and $\kappa_{alb}$, respectively. The dynamics of those uninfected adults are influenced by the adult carrying capacity ($K_{ad}$). Uninfected adults could acquire the dengue virus when biting infected humans ($I$), with transmission rate as $\zeta_{ae}$ ($p_DBR_{ae}$) and $\zeta_{alb}$ ($p_DBR_{alb}$) representing the probability of infection multiplied by the bitting rate for *Ae. aegypti* and *Ae. albopictus*, respectively. Once infected, these adults transitioned to the infected state ($Y_{ae}$ and $Y_{alb}$ for each species). The dynamics of infected adults were further influenced by adult mortality rates ($\mu_{ae}$ and $\mu_{alb}$).

Table 1: Parameter values for the dengue transmission model with and without temperature dependence.

| Parameter | Symbol | Value (Without Temperature) | Value (With Temperature) | Units | Reference |
|---|---|---|---|---|---|
| Adult mortality rate (Ae. aegypti) | $\mu_{Aae}$ | 1/10 | $(0.05 \cdot (T - 25.5))^2 + 0.015$ | days$^{-1}$ | [21, 22] |
| Adult mortality rate (Ae. albopictus) | $\mu_{Aalb}$ | 1/12 | $(0.05 \cdot (T - 25.5))^2 + 0.01$ | days$^{-1}$ | [21, 22] |
| Biting rate (Ae. aegypti) | $BR_{ae}$ | 0.470 | $\frac{1}{1.1\sqrt{2\pi}} e^{\left(-\frac{(T-33)^2}{2 \cdot (1.1)^2}\right)}$ | $\frac{Humans}{Mosquitoes \cdot Time}$ | [7, 23–25] |
| Biting rate (Ae. albopictus) | $BR_{alb}$ | 0.350 | $\frac{1}{1.4\sqrt{2\pi}} e^{\left(-\frac{(T-33)^2}{2 \cdot (1.4)^2}\right)}$ | $\frac{Humans}{Mosquitoes \cdot Time}$ | [23, 24] |
| Larval Carrying capacity | $K$ | 3000 | 3000 | Individuals | Estimated |
| Adult Carrying capacity | $K_{ad}$ | 10000 | 10000 | Individuals | Estimated |
| dengue transmission probability | $P_D$ | 0.700 | $4.8 \cdot \frac{1}{2.4\sqrt{2\pi}} e^{\left(-\frac{(T-30)^2}{2 \cdot (2.4)^2}\right)}$ | Dimensionless | Estimated |
| Human incubation rate | $\theta$ | 1/10 | 1/10 | days$^{-1}$ | [26] |
| Human recovery rate | $\gamma$ | 1/7 | 1/7 | days$^{-1}$ | [26] |
| Larval comp. coeff. (Ae. aegypti) | $\omega_{ae}$ | 0 to 1 | 0 to 1 | Dimensionless | Fixed |
| Larval comp. coeff. (Ae. albopictus) | $\omega_{alb}$ | 0 to 1 | 0 to 1 | Dimensionless | Fixed |
| Larval develop. rate (Ae. aegypti) | $\kappa_{ae}$ | 0.170 | $\frac{1}{2.7\sqrt{2\pi}} e^{\left(-\frac{(T-33)^2}{2 \cdot (2.7)^2}\right)}$ | days$^{-1}$ | [24, 28] |
| Larval develop. rate (Ae. albopictus) | $\kappa_{alb}$ | 0.167 | $\frac{1}{2.4\sqrt{2\pi}} e^{\left(-\frac{(T-33)^2}{2 \cdot (2.4)^2}\right)}$ | days$^{-1}$ | [24, 29] |
| Larval mortality rate (Ae. aegypti) | $\mu_{Lae}$ | 0.090 | $(0.05 \cdot (T - 25.5))^2 + 0.09$ | days$^{-1}$ | [22, 28] |
| Larval mortality rate (Ae. albopictus) | $\mu_{Lalb}$ | 0.050 | $(0.05 \cdot (T - 25.5))^2 + 0.05$ | days$^{-1}$ | Estimated [22] |
| Larval growth rate (Ae. aegypti) | $\beta_{ae}$ | 0.340 | $70 \cdot \frac{1}{3.5\sqrt{2\pi}} e^{\left(-\frac{(T-27)^2}{2 \cdot (3.5)^2}\right)}$ | days$^{-1}$ | [24, 30, 31] |
| Larval growth rate (Ae. albopictus) | $\beta_{alb}$ | 0.253 | $64 \cdot \frac{1}{3.5\sqrt{2\pi}} e^{\left(-\frac{(T-27)^2}{2 \cdot (3.5)^2}\right)}$ | days$^{-1}$ | [24, 29] |
| Temperature | $T$ | - | $2.7 \cdot \sin(-0.04 \cdot T) + 29.8$ | °C | Estimated |

Note: many of the temperature-dependent functions were modeled as Gaussian functions.

$$\dot{L}_{ae} = \beta_{ae}L_{ae}\left(1 - \frac{L_{ae} + \omega_{alb}L_{alb}}{K}\right) - \kappa_{ae}L_{ae} - \mu_{Lae}L_{ae} \tag{1}$$

$$\dot{L}_{alb} = \beta_{alb}L_{alb}\left(1 - \frac{L_{alb} + \omega_{ae}L_{ae}}{K}\right) - \kappa_{alb}L_{alb} - \mu_{Lalb}L_{alb} \tag{2}$$

$$\dot{X}_{ae} = \kappa_{ae}L_{ae}\left(1 - \frac{X_{ae}}{K_{ad}}\right) - \zeta_{ae}\frac{X_{ae}I}{N_h} - \mu_{ae}X_{ae} \tag{3}$$

$$\dot{X}_{alb} = \kappa_{alb}L_{alb}\left(1 - \frac{X_{alb}}{K_{ad}}\right) - \zeta_{alb}\frac{X_{alb}I}{N_h} - \mu_{alb}X_{alb} \tag{4}$$

$$\dot{Y}_{ae} = \zeta_{ae}\frac{X_{ae}I}{N_h}\left(1 - \frac{Y_{ae}}{K_{ad}}\right) - \mu_{ae}Y_{ae} \tag{5}$$

$$\dot{Y}_{alb} = \zeta_{alb}\frac{X_{alb}I}{N_h}\left(1 - \frac{Y_{alb}}{K_{ad}}\right) - \mu_{alb}Y_{alb} \tag{6}$$

The human population dynamics were sub-divided into four compartments: susceptible (*S*), exposed (*E*), infected (*I*), and recovered (*R*).

$$\dot{S} = -\lambda S \tag{7}$$

$$\dot{E} = \lambda S - \theta E \tag{8}$$

$$\dot{I} = \theta E - \gamma I \tag{9}$$

$$\dot{R} = \gamma I \tag{10}$$

Susceptible individuals (*S*) could acquire the infection through bites from infected mosquitoes. The forces of infection ($\lambda$) were calculated as [32]:

$$\lambda_h = \frac{p_D(BR_{ae}Y_{ae} + BR_{alb}Y_{alb})}{N_h} \tag{11}$$

$$\lambda_{ae} = \frac{\zeta_{ae}I}{N_h} \tag{12}$$

$$\lambda_{alb} = \frac{\zeta_{alb}I}{N_h} \tag{13}$$

The total human population is represented as $N_h = S + E + I + R$ is total human population, and $p_D$ is the dengue transmissibility from mosquitoes to humans. Exposed individuals ($E$) progressed to the infected compartment ($I$) at a rate $\theta$, while infected individuals recovered at a rate $\gamma$, transitioning to the recovered compartment ($R$). The model employs a constant human population size ($N_h$ = 1,000) for two reasons. First, the assumption that the short-term simulation horizon (1-2 years) renders natural birth/death processes epidemiologically negligible compared to dengue transmission dynamics in a locality of a city. Second, maintaining demographic stability allows focused analysis of temperature-driven vector competition effects without introducing confounding parameters.

In the temperature-dependent model, we incorporated thermal sensitivity into six key parameters: (1) biting rates, (2) larval development rates, (3) larval growth rates, (4) larval mortality, (5) adult mortality, and (6) dengue transmission probability (see Table 1). Each parameter follows an approximation for experimentally validated temperature-response functions that capture non-linear thermal optima and limits for both *Aedes* species.

## 2.2 Invasion analysis

We assessed the invasion potential of *Ae. albopictus* through the theoretical framework of fitness as defined by the dominant Lyapunov exponent ($\rho$) [33], which quantifies the asymptotic exponential growth rate of a population in a given environment. We implemented this ecological interpretation as:

$$\rho = \frac{\log(N_f/N_i)}{t} \tag{14}$$

Where $N_i$ and $N_f$ represent the initial and final population densities of *Ae. albopictus* (the invader), awhile $t$ indicates a simulation period of 1,000 days. We calculated this metric across a range of larval competition coefficients ($\omega_{ae}, \omega_{alb} \in [0.1, 1.0]$) under two scenarios: (1) a temperature-independent baseline model and (2) a temperature-dependent model where development, mortality, fecundity, and biting rates responded to thermal conditions. The initial conditions reflected a system dominated by *Ae. aegypti* ($L_{ae}$ = 1000, $L_{alb}$ = 10) to assess invasion potential against an established resident population. Following theoretical expectations [33, 34], we classified parameter combinations with $\rho > 0$ as permitting invasion (indicating positive fitness), whereas $\rho < 0$ denoted competitive exclusion. Cases where $\rho \approx 0$ were interpreted as neutral equilibria, representing a state in which neither species has a competitive advantage and where small perturbations in species abundances neither grow nor decay over time—conditions typically referred to as "evolutionary equilibria".

## 2.3 Coexistence analysis

We simulated both temperature-independent and temperature-dependent models to describe the dynamics of larval and adult populations of *Ae. aegypti* and *Ae. albopictus*. Our goal was to explore species coexistence under different scenarios of larval interspecific competition. The analysis concentrated on assessing the effects of larval competition coefficients, denoted by $\omega_{ae}$ and $\omega_{alb}$, which control the competitive impact that each species has on the other.

We systematically varied the parameters across a defined range to create a grid of scenarios and solved the system of differential equations for each pair of parameters. For each combination, we calculated the final abundances of the larval stages and quantified the relative proportions of each species. This data allowed us to identify the conditions under which one species dominates, both species coexist, or competitive exclusion occurs. We summarized these outcomes using violin plots to illustrate the distribution of larval proportions across all simulated scenarios.

We conducted statistical comparisons between temperature-independent and temperature-dependent models using the Wilcoxon rank-sum test (also known as the Mann-Whitney U test). This non-parametric test was selected due to the non-normal distribution of relative abundance data and the presence of heteroscedasticity across parameter combinations, which violated the assumptions required for parametric tests such as the t-test. The objective of the statistical analysis was to determine whether temperature effects significantly altered the competitive balance between species by comparing the distributions of relative abundances under each modeling scenario. We evaluated the null hypothesis that the median relative abundances of each species were equivalent between temperature-independent and temperature-dependent conditions, with significance assessed at $\alpha = 0.05$.

## 2.4 Effects on dengue dynamics

We evaluated how interspecific larval competition affects dengue transmission dynamics under three ecologically relevant initial conditions: (1) equal species abundance, (2) *Ae. aegypti* dominance, reflecting typical field observations, and (3) *Ae. albopictus* dominance. For each condition, we simulated both temperature-dependent and temperature-independent models to compare thermal effects on transmission outcomes. Human infection peaks were quantified relative to the larval competition coefficients, and the results were visualized using heatmaps to illustrate how varying competitive pressures shape transmission intensity across scenarios.

We calculated the mathematical expression of basic reproduction number ($R_0$) analytically using the Next Generation Matrix (NGM) method. This procedure involved: (1) defining the infection and transition matrices to represent all transmission pathways between human and mosquito compartments, (2) computing the spectral radius (dominant eigenvalue) of the resulting matrix product, and (3) deriving a closedform expression for $R_0$. Since $R_0$ represents a static measure, we complemented the analysis with the effective reproduction number $R_t$, calculated in random time step during the growth phase of the human infectious dynamic, which captures the dynamic effects of larval competition throughout the simulation on transmission potential. All graphical analyses of transmission dynamics, including heatmaps and scenario comparisons, were based on $R_t$ values obtained from simulated trajectories under both temperature-independent and temperature-dependent conditions.

This framework combines analytical and numerical approaches to provide a comprehensive analysis. The analytical calculation of $R_0$ serves as a baseline measure for the intrinsic transmission potential of the system, taking into account various ecological and thermal configurations, while remaining independent of transient dynamics. Concurrently, numerical simulations of $R_t$ allow us to observe temporal changes in transmission potential, which are influenced by shifts in vector abundance due to larval competition and temperature effects. Human infection peaks were analyzed alongside $R_t$ trajectories to assess how fluctuations in vector populations translated into changes in epidemiological risk over time. By implementing $R_0$, $R_t$, and infection peaks, we ensured that the static theoretical estimates and the dynamic simulations could be interpreted together, enabling a consistent evaluation of the influence of larval competition and temperature on dengue transmission.

## 3 Results

### 3.1 Invasion analysis

The invasion dynamics revealed distinct ecological regimes across the larval competition parameter space ($\omega_{ae}$, $\omega_{alb} \in [0,1.0]$). Figure 2 displays the results for both temperature-independent (left panel) and temperature-dependent (right panel) models.

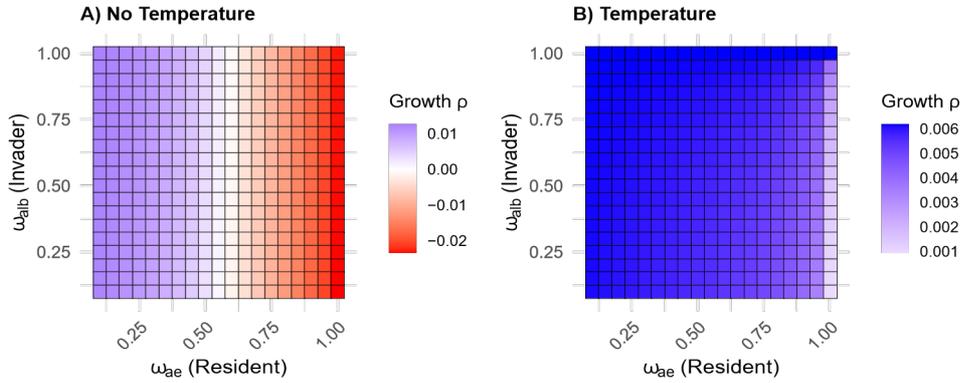

**Fig. 2**: Invasion analysis for *Ae. albopictus* in a system dominated by *Ae. aegypti*. A) Temperature-independent model showing three distinct zones: 1) successful invasion ($\rho > 0$) when $\omega_{ae} < 0.47$, 2) neutral equilibrium ($\rho \approx 0$) for $0.47 \leq \omega_{ae} \leq 0.60$, and 3) competitive exclusion ($\rho < 0$) at $\omega_{ae} > 0.60$. B) Temperature-dependent model exhibiting expanded invasion potential ($\rho > 0$ for $\omega_{ae} \leq 0.75$) with only marginal exclusion ($\rho \approx 0$) in high competition scenarios ($\omega_{ae} > 0.75$). Color gradients represent Lyapunov exponent ($\rho$) values quantifying invasion fitness, with warmer colors indicating higher establishment probability.

In the temperature-independent model, three distinct characteristic regimes emerged with Lyapunov coefficients ranging from -0.02 to 0.01. Firstly, *Ae. albopictus* exhibited positive Lyapunov coefficients ($\rho > 0$) when the larval competition coefficient of *Ae. aegypti* ($\omega_{ae}$) ranged from 0 to approximately 0.47 regardless of its own competition coefficient ($\omega_{alb}$). This indicates favorable conditions for invasion, with a higher potential for *Ae. albopictus*. Secondly, a narrow transition zone between $\omega_{ae}$ values of 0.47 and 0.60 exhibited Lyapunov coefficients at or near zero ($\rho \approx 0$) for all $\omega_{alb}$ values. This situation corresponds to a neutral equilibrium and could represent evolutionary bifurcation boundaries. At this point, the long-term dynamics may shift toward coexistence, dominance by one species, or exclusion of one species, depending on other ecological or environmental factors. Lastly, for $\omega_{ae}$ values exceeding 0.60, negative Lyapunov coefficients ($\rho < 0$) were found for *Ae. albopictus* across all $\omega_{alb}$ values, suggesting competitive exclusion by *Ae. aegypti*.

The temperature-dependent model exhibited distinct dynamics, with Lyapunov coefficients ranging from 0.001 to 0.006. These coefficients consistently showed positive values across all parameter combinations. The model indicated *Ae. albopictus* has a strong invasion capability under varying thermal conditions, as evidenced by the positive Lyapunov coefficients observed in the majority of the $\omega_{ae}$ parameter space. Invasion success persisted even at high competition levels ($\omega_{ae} = 0.75$), demonstrating a substantial increase in invasion potential compared to temperature-independent conditions. A triangular zone was identified for $\omega_{ae}$ values between 0.75 and 1.0, where the Lyapunov coefficient for *Ae. albopictus* decreased toward the lower end of the range, suggesting that competitive pressure from *Ae. aegypti* reached an extreme level of competition asymmetry, yet *Ae. albopictus* still maintained a diminished, but not fully suppressed, potential for invasion.

A comparison between models revealed that the temperature-dependent model has a wider parameter space with positive Lyapunov coefficients compared to the temperature-independent model. This suggests that temperature influences the dynamics of invasion. In the temperature-dependent scenario the area with negative coefficients was nearly eliminated. This indicates that the thermal modulation of life history parameters significantly enhances the ability of *Ae. albopictus* to establish itself in areas dominated by *Ae. aegypti*.

## 3.2 Coexistence analysis

The coexistence analysis measured the relative abundances of both mosquito species under contrasting thermal scenarios across multiple parameter combinations (Figure 3). The violin plot displays the results for temperature-independent (left panel) and temperature-dependent (right panel) models.

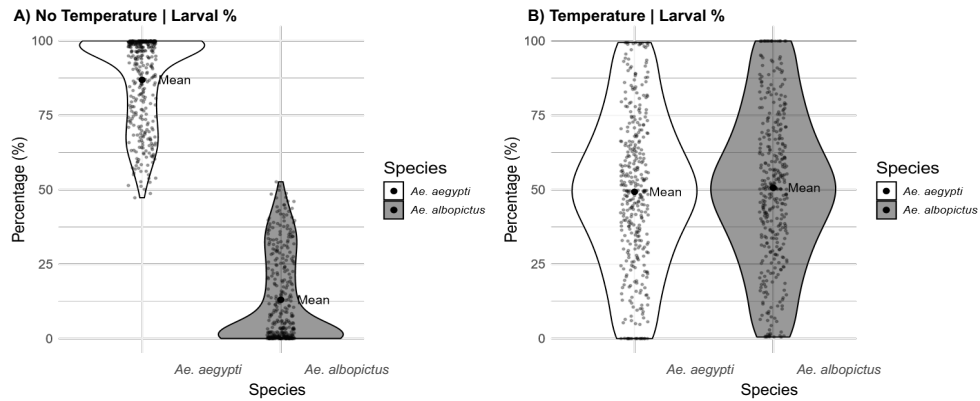

**Fig. 3**: Coexistence analysis of *Aedes* mosquitoes. A) The analysis showed that in the temperature-independent model, *Ae. aegypti* dominated with 87.5% larval abundance in the system, while *Ae. albopictus* had only 12.5%. B) In contrast, the temperature-dependent model displayed a more balanced distribution, with both species reaching around 50% larval abundance, enabling coexistence.

In the temperature-independent model, *Ae. aegypti* displayed a clear competitive dominance over *Ae. albopictus* across the entire parameter space examined. The species distribution was notably asymmetric, *Ae. aegypti* achieved a mean relative abundance of 87.5% (range: 50-100%), whereas *Ae. albopictus* was consistently restricted to 12.5% (range: 0-50%). The violin plot for *Ae. aegypti* showed a distribution heavily skewed toward higher abundance values, with most observations concentrated above 70%. In contrast, the distribution for *Ae. albopictus* was compressed toward lower values. This competitive imbalance was statistically significant (Wilcoxon rank-sum test: $W = 130{,}305$, $p < 2.2 \times 10^{-16}$), confirming substantial differences in species performance under temperature-independent conditions and indicating complete competitive hierarchy between species.

The temperature-dependent model revealed a shift in the competitive balance, resulting in a greater potential for the coexistence of species. This wide spread indicates substantial fluctuations around the mean, underscoring that coexistence occurred under a broad set of competitive conditions. The violin plots displayed substantial overlap between species, with both distributions showing similar shapes and peak densities centered around the 50% mark. Unlike the temperature-independent scenario, neither species displayed pronounced skewness; instead, their distributions were more symmetric around the central tendency. The interquartile ranges for both species were comparable, indicating similar variability in competitive outcomes. Statistical analysis confirmed the absence of significant competitive advantage for either species (Wilcoxon rank-sum test: $W = 63{,}097$, $p = 0.462$), demonstrating that temperature effects effectively neutralized the competitive hierarchy observed in the temperature-independent scenario and enabled balanced population dynamics between both mosquito species.

## 3.3 Effects on dengue transmission

The infection analysis examined transmission dynamics under *Ae. aegypti* dominance scenarios, comparing temperature-independent and temperature-dependent conditions (Figure 4). In the temperature-independent scenario (left panel), infection values remained constrained within a narrow range of 156-168 infected individuals. Peak infections consistently occurred under low competition scenarios, with the highest transmission potential observed when both larval competition coefficients ($\omega_{ae}$ and $\omega_{alb}$) approached minimal values near 0. Conversely, the lowest infection values emerged when *Ae. albopictus* competition coefficients ($\omega_{alb}$) reached maximum values (0.75-1.00), demonstrating the constraining effect of interspecific competition on transmission dynamics.

The temperature-dependent scenario (right panel) exhibited substantially elevated and more variable infection patterns compared to the temperature-independent condition. Infection values expanded to 195-220 infected individuals, representing a 25-30% increase from the temperature-independent scenario. Under optimal low-competition conditions, peak infections exceeded 210, with the highest values concentrated in areas where both larval competition coefficients remained low. Although there was an overall increase in transmission potential, the spatial distribution of infection peaks followed similar patterns. Minimum infection values still occurred when competition from *Ae.*

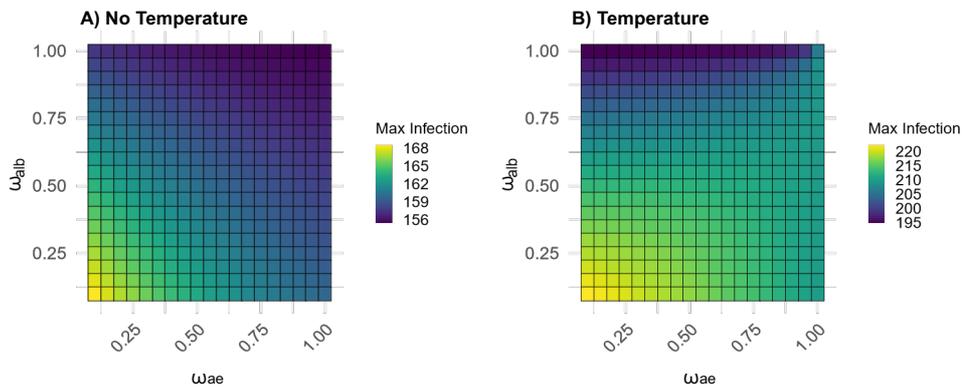

**Fig. 4**: Dengue transmission dynamics under *Aedes aegypti* dominance across temperature conditions. Heatmaps showing the number of infected individuals as a function of larval competition coefficients for both species ($\omega_{ae}$ and $\omega_{alb}$) when *Ae. aegypti* has higher initial abundance. Panel A: Temperature-independent scenario showing infection values constrained to 156-168 infected individuals, with peak transmission occurring at minimal competition coefficients (both $\omega_{ae}$ and $\omega_{alb}$ near 0) and lowest values when *Ae. albopictus* competition reaches maximum (0.75-1.00). Panel B: Temperature-dependent scenario exhibiting elevated infection patterns (195-220 infected individuals), representing a 25-30% increase compared to temperature-independent conditions. Peak infections exceed 210 under optimal low-competition conditions, while maintaining similar spatial distribution patterns with maximum transmission concentrated in low-competition regions and minimum values occurring under high *Ae. albopictus*

competition. Color scales represent the number of infected individuals, with warmer colors indicating higher transmission potential.

*albopictus* reached maximum coefficients, though the absolute infection levels were consistently elevated compared to temperature-independent conditions.

Additional scenarios examining equal initial abundances and *Ae. albopictus* dominance revealed complementary patterns (see Supplementary Figure 1). Under conditions with equal initial abundances, temperature-independent conditions resulted in infection values of 150-170 infected individuals, while temperature-dependent conditions elevated these values to 190-220 infected individuals. When *Ae. albopictus* achieved greater initial abundance, infection values were compressed to 140-170 infected individuals under temperature-independent conditions and 190-215 infected individuals under temperature-dependent conditions. This represented the lowest transmission potential across all scenarios within their respective thermal conditions.

### 3.3.1 $R_0$ analysis

The basic reproduction number ($R_0$) for the model was calculated using the next generation matrix method (see Supplementary Calculations, subsection 6.2):

$$R_0 = p_D \sqrt{\frac{BR_{ae}^2 \mu_{Aalb} x_{ae} + BR_{alb}^2 \mu_{Aae} x_{alb}}{N_h \gamma \mu_{Aae} \mu_{Aalb}}} \tag{15}$$

The $R_0$ expression indicates that transmission potential arises from the combined contributions of infected adult mosquitoes of both species, with key drivers being the biting rates ($BR_{ae}$ and $BR_{alb}$), transmission probabilities ($\zeta_{ae}$ and $\zeta_{alb}$), and adult mortality rates ($\mu_{Aae}$ and $\mu_{Aalb}$). While larval stages are not explicitly represented in this formula, larval competition regulates the abundance of susceptible adults ($X_{ae}$ and $X_{alb}$), indirectly influencing $R_0$ in a natural escenario by controlling the size of the vector population available for transmission.

In the temperature-independent model, $R_t$ values ranged from 2.05 to 2.20 (Figure 5). The highest $R_t$ values occurred when both larval competition coefficients ($\omega_{ae}$ and $\omega_{alb}$) were minimal, consistent with the $R_0$ expectation that reduced competition increases vector abundance and thus transmission potential. As competition intensified, $R_t$ declined to a minimum of 2.05 when both coefficients reached their maximum values ($\omega_{ae} = \omega_{alb} = 1$), and reflects how stronger interspecific competition suppresses mosquito populations and limits transmission. When temperature effects were incorporated into the model, $R_t$ values shifted to a range of 1.00 to 1.80, with elevated values limited under lower larval competition coefficients. The lowest $R_t$ value of 1.00 occurred when *Ae. albopictus* reached maximum larval competition, suggesting that temperature variations can intensify the negative impact of competitive dominance on transmission potential. The overall reduction in $R_t$ values under thermal influence reflects temperature-mediated changes to vector survival rates and development times, which impose additional constraints on transmission capacity, particularly under conditions of intense larval competition.

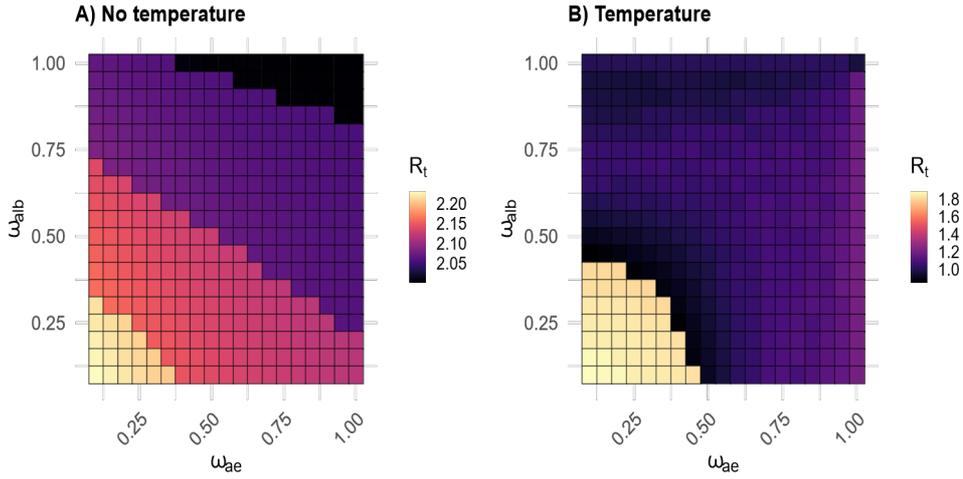

**Fig. 5**: Effect of Competition and Temperature on $R_t$ in *Aedes* Mosquitoes: Without temperature effects, $R_t$ ranged from 2.05 to 2.20, peaking under minimal larval competition. With temperature, $R_t$ dropped to 1.00 to 1.80, with the lowest values under intense competition. High $R_t$ persisted in low-competition, warm conditions, highlighting *Ae.aegypti*'s strong transmission potential.

## 4 Discussion

Our results demonstrate that temperature-mediated competition profoundly alters the invasion potential. Under constant temperature conditions, *Ae. albopictus* invasion thresholds expands from $\omega_{ae} < 0.47$ under constant temperature conditions to $\omega_{ae} = 0.75$ when temperature dependence is incorporated. Crucially, without these temperature variations, the observed large-scale invasion of *Ae. albopictus* would not occur, as thermal fluctuations create temporally variable conditions that favor its competitive establishment. This is quantitatively reflected in the Lyapunov coefficient shifts from mixed values (−0.02 to 0.01) to exclusively positive ranges (0.001 to 0.006), representing an order-of-magnitude change in invasion stability. The temperature-dependent advantage emerges through multiple mechanisms: enhanced development rates, increased adult survival, and higher fecundity for *Ae. albopictus*, coupled with modified interaction strengths between the species. These findings align with field observations documenting *Ae. albopictus*-driven displacement of *Ae. aegypti* populations [35], particularly in resource-limited environments where its superior environmental tolerance provides competitive advantages [36]. The consistent positivity of invasion metrics under temperature fluctuations suggests that climate change and global warming may further facilitate *Ae. albopictus* range expansion into traditionally *Ae. aegypti*-dominated regions, as evidenced by comparative studies of their thermal performance [37, 38].

The temperature-dependent model demonstrates a key transition from competitive exclusion (87.5% *Ae. aegypti* dominance, $p < 2.2 \times 10^{-16}$) to stable coexistence (50% abundance each, $p = 0.462$). This indicates that thermal variability can help balance interspecific competition. The effect is more pronounced than in constant-temperature models because annual temperature fluctuations allow for flexible resource partitioning, which static environments cannot support. Field studies corroborate these modeled dynamics, showing stable coexistence in regions with seasonal temperature fluctuations [39, 40]. While *Ae. albopictus* benefits from temperature-mediated larval competition advantages [41], its coexistence with *Ae. aegypti* can also be modulated by additional mechanisms reported in previous studies, including cross-species mating interference [42] and differential sensitivity to desiccation stress [43]. Importantly, the model predicts that climate change may alter current coexistence dynamics by unevenly impacting the thermal preferences of each species. This could reshape vector community composition in areas where these species are endemic. The persistence of both species across varying environmental conditions suggests that complete competitive exclusion is uncommon in natural systems experiencing temperature variability. This aligns with observations of sympatric populations in multiple biogeographic regions and the ability of *Ae. albopictus* to persist in places where *Ae. aegypti* cannot [15, 42].

Temperature effects revealed complex interactions that challenge some of our initial predictions. The increased temperature dependence led to a 25–30% rise in infection rates compared to scenarios with constant temperatures, indicating that the thermal enhancement of vectorial capacity outweighs the suppression caused by competition. The bidirectional effect on transmission is more physiologically realistic in the temperature-dependent model because vectors respond to seasonal variations rather than being restricted by constant temperatures. While *Ae. aegypti* remains its role as primary vector [44], the temperature-modulated competition significantly constrains transmission at high larval densities [45–48]. This complex interplay suggests that climate change may heighten the risks of dengue through two mechanisms: the thermal optimization of vector traits and competition-driven population dynamics. Notably, the model captures the epidemiologically significant transition where *Ae. albopictus*dominated systems show attenuated transmission compared to *Ae. aegypti* scenarios [44, 49], yet still maintain substantial outbreak potential in temperate regions [50, 51]. The situation where both vectors exist in similar abundances is particularly important from an epidemiological perspective. This balanced coexistence allows disease transmission to continue under a broader range of environmental conditions than if either species were present alone. The increased adaptability of temperature-dependent interactions highlights the need to consider climatic variability when forecasting disease transmission in the context of global change scenarios.

Our hypothesis regarding bidirectional competition effects was strongly supported by the $R_0$ and $R_t$ analysis and infection dynamics. The lowest infection values consistently appeared when both species faced high larval competition. Specifically, when competition was at its maximum in temperature-independent conditions, Rt values declined from 2.20 to 2.05 under maximum competition in temperature-independent conditions, and from 1.80 to 1.00 when temperature effects were incorporated. This pattern confirms that interspecific competition constrains vector population sizes and reduces transmission capacity. Conversely, the highest transmission peaks occurred under symmetric low competition ($\omega_{ae}$ and $\omega_{alb}$ near 0), supporting our prediction that reduced competition enhances vector abundance and amplifies transmission potential. Additionally, the persistence of relatively high Rt values in low-competition scenarios under temperature-dependent conditions highlights *Ae. aegypti*'s ability to sustain high transmission peaks in thermally favorable environments, consistent with its role as a primary epidemic vector [44].

## 5 Conlussions

The study's implications for public health strategies are significant, particularly in regions like Colombia where both species coexist. Vector control programs must consider not only individual species abundance but also their competitive interactions and the effects of temperature variation. Such dynamics are especially relevant under ongoing and future climate change, as shifts in temperature regimes may favor *Aedes albopictus* persistence in rural and urban areas where *Aedes aegypti* has historically predominated.

Despite these advances, several limitations warrant future investigation. The model's focus on temperature as the primary environmental variable overlooks other factors such as humidity, precipitation, and resource availability that may influence vector dynamics. The relative scarcity of parameterization data for *Ae. albopictus* compared to *Ae. aegypti* may affect prediction accuracy. Future studies should aim to collect more comprehensive field data on *Ae. albopictus*, investigate additional environmental variables, and explore how climate change might modify the competitive dynamics between these species. Vector surveillance programs should include regular monitoring of the populations and relative abundances of both species, particularly in areas where their distributions overlap. This approach will help develop more effective strategies for mitigating dengue transmission where both vectors coexist.

## Declarations

### Ethics approval and consent to participate

Not applicable.

### Consent for publication

Not applicable.

### Availability of data and materials

The computational codes developed and used for this study are available in the GitHub repository https://github.com/savch1102/Dengue-temperature-code-files.git.

### Competing interests

The authors declare that they have no competing interests.

## Funding

This study was funded by Grant NFS DMS-2327814. The funder had no role in the study design, data analysis, decision to publish, or preparation of the manuscript.

## Authors' contributions

SV was responsible for programming and implementing the model in R and Python. Together, SV and MS designed and conducted the computational results analyses, interpreting the results and ensuring the robustness of the study. Both authors jointly wrote the manuscript, engaging in continuous discussions throughout the project, providing critical feedback at every stage, and approving the final version of the manuscript.

## Acknowledgements

We thank Grant NFS DMS-2327814 for financial support, Carlos Bravo for his guidance throughout the project, and the group of Biología Matemática y Computacional de la Universidad de Los Andes (BIOMAC) for providing an excellent research environment and constructive feedback enriched with strong human values.

# Supplementary Information

Santiago Villamil & Mauricio Santos Vega

Universidad de Los Andes

---

# 1 Calculation of the Basic Reproduction Number ($R_0$)

In this section, we detail the calculation of the basic reproduction number ($R_0$) for the mosquito-human competition model. The state variables and parameters are defined as follows:

State Variables

- $S$: Susceptible humans.
- $E$: Exposed humans.
- $I$: Infected humans.
- $Y_{ae}$: Infected *Aedes aegypti* mosquitoes.
- $Y_{alb}$: Infected *Aedes albopictus* mosquitoes.

Parameters

- $N_h$: Total human population.
- $p_D$: Probability of disease transmission per bite.
- $BR_{ae}$, $BR_{alb}$: Mosquito biting rates for *Aedes aegypti* and *Aedes albopictus*, respectively.
- $\zeta_{ae}$, $\zeta_{alb}$: Transmission efficiency from human to mosquito
- $X_{ae}$, $X_{alb}$: Mosquito population sizes for *Aedes aegypti* and *Aedes albopictus*.
- $\theta_h$: Rate of progression from exposed to infected in humans.
- $\gamma_h$: Recovery rate of infected humans.
- $\mu_{Aae}$, $\mu_{Aalb}$: Mortality rates of infected *Aedes aegypti* and *Aedes albopictus*.

Next-Generation Matrix

The infection dynamics are governed by the following vectors of new infections (**F**) and transitions (**V**) (see subsection ??):

$$\mathbf{F} = \begin{pmatrix} \frac{p_D(BR_{ae}Y_{ae}+BR_{alb}Y_{alb})}{N_h}S \\ 0 \\ \frac{\zeta_{ae}X_{ae}I}{N_h}\left(1 - \frac{Y_{ae}}{K_{ad}}\right) \\ \frac{\zeta_{alb}X_{alb}I}{N_h}\left(1 - \frac{Y_{alb}}{K_{alb}}\right) \end{pmatrix},$$

$$\mathbf{V} = \begin{pmatrix} \theta_h E \\ -\theta_h E + \gamma_h I \\ \mu_{Aae}Y_{ae} \\ \mu_{Aalb}Y_{alb} \end{pmatrix}.$$

The Jacobian matrices of **F** and **V** evaluated at the disease-free equilibrium yield the next-generation matrix:

$$\mathcal{K} = \begin{bmatrix} 0 & 0 & \frac{BR_{ae}p_D}{\mu_{Aae}} & \frac{BR_{alb}p_D}{\mu_{Aalb}} \\ 0 & 0 & 0 & 0 \\ \frac{BR_{ae}p_D x_{ae}}{N_h\gamma} & \frac{BR_{ae}p_D x_{ae}}{N_h\gamma} & 0 & 0 \\ \frac{BR_{alb}p_D x_{alb}}{N_h\gamma} & \frac{BR_{alb}p_D x_{alb}}{N_h\gamma} & 0 & 0 \end{bmatrix}$$

Basic Reproduction Number ($R_0$)

The basic reproduction number is the spectral radius (dominant eigenvalue) of the nextgeneration matrix:

$$R_0 = p_D\sqrt{\frac{BR_{ae}^2\mu_{Aalb}x_{ae} + BR_{alb}^2\mu_{Aae}x_{alb}}{N_h\gamma\mu_{Aae}\mu_{Aalb}}}$$

Note: matrix algebra was calculated with Python.

# 2 Effects on dengue transmission

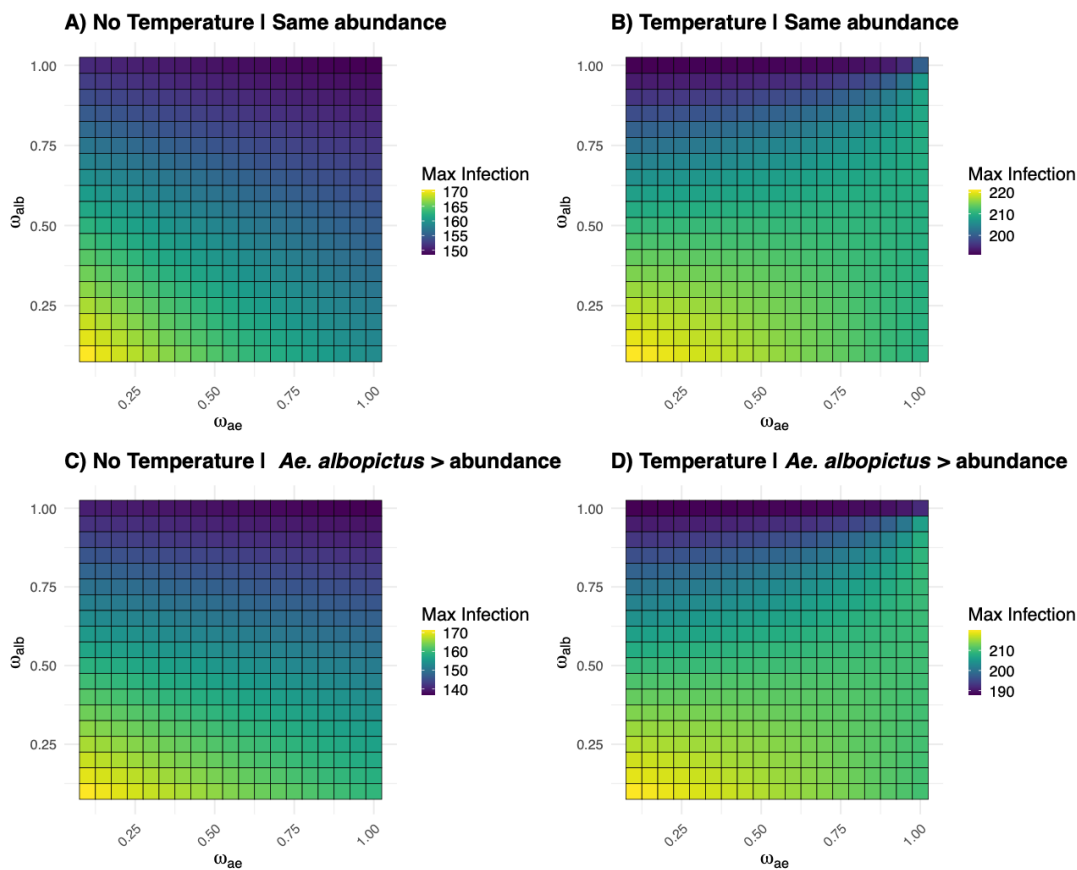

**Fig. S1**: Dengue transmission dynamics under equal abundances and *Aedes albopictus* dominance scenarios. Heat maps comparing infection patterns across initial abundance compositions and temperature conditions. Panels A-B: Equal initial abundances showing infection values of 150-170 individuals under temperature-independent conditions (Panel A) and 190-220 individuals under temperature-dependent conditions (Panel B). Panels C-D: *Ae. Albopictus* dominance scenarios displaying the lowest transmission potential with infection values of 140-170 individuals under temperature-independent conditions (Panel C) and 190-215 individuals under temperature-dependent conditions (Panel D). Across all scenarios, peak infections consistently occur when both larval competition coefficients ($\omega_{ae}$ and $\omega_{alb}$) approach minimal values near 0, while maximum *Ae. albopictus* competition coefficients (0.75-1.00) produce the lowest infection values. Temperature-dependent conditions consistently elevate transmission potential by 25-30% compared to their temperature-independent counterparts, regardless of initial abundance composition. Color scales represent the number of infected individuals, with spatial patterns demonstrating the constraining effect of interspecific competition on dengue transmission dynamics.